\documentclass[twocolumn]{aastex701}
\usepackage{amsmath} 
\usepackage{tabularx,booktabs} 
\usepackage{gensymb} 
\usepackage{subcaption} 
\usepackage{xcolor}
\usepackage{float} 
\usepackage{braket}

\begin{document}

\title{Beyond the Impulsive Approximation: The Dynamics and Radio Emission of Tidal Disruption Jets}

\author[orcid=0000-0002-3137-4633]{Fabio De Colle}
\affiliation{Instituto de Ciencias Nucleares, Universidad Nacional Autónoma de México, Apartado Postal 70-543, Ciudad de México 04510, Mexico}
\email{fabio.decolle@nucleares.unam.mx}

\author[orcid=0000-0003-2558-3102]{Enrico Ramirez-Ruiz} 
\affiliation{Department of Astronomy and Astrophysics, University of California, Santa Cruz, CA 95064, USA}
\email{enrico@ucolick.org}

\begin{abstract}
Radio emission from relativistic tidal disruption events (TDEs) is commonly interpreted using the Blandford--McKee solution for an impulsive relativistic blast wave. Unlike gamma-ray bursts, however, TDE jets are powered by fallback accretion, injecting energy over weeks to months. We investigate how this sustained energy injection modifies the dynamics and radio emission of relativistic TDE jets using one- and two-dimensional relativistic hydrodynamic simulations coupled with synchrotron radiation calculations. For a fixed total energy, fallback-powered jets drive slower forward shocks, producing delayed and systematically fainter radio emission. We show that the impulsive approximation is valid only when the duration of energy injection is less than approximately ten percent of the deceleration time of an equivalent Blandford--McKee blast wave, a condition that is generally not satisfied for relativistic TDEs. As a result, continuously powered outflows are significantly less beamed than impulsive explosions, leading to a strong suppression of off-axis radio emission. These results demonstrate that the radio evolution of relativistic TDEs retains memory of the central engine and cannot, in general, be described by an impulsive solution.  By substantially reducing the expected off-axis radio emission, sustained fallback-powered jets imply that relativistic jet production in tidal disruption events may be considerably more common than current radio observations suggest.
\end{abstract}


\section{Introduction}
Tidal disruption events (TDEs) occur when a star passes sufficiently close to a supermassive black hole to be torn apart by tidal forces, producing a luminous flare powered by the return and subsequent accretion of stellar debris \citep{Hills1975,Rees1988,Phinney1989,StrubbeQuataert2009,GuillochonRamirezRuiz2013,Mockler2019}. In addition to their thermal optical, ultraviolet, and X-ray emission \citep{Auchettl2017, Gezari2021}, many TDEs produce outflows that interact with the circumnuclear medium, giving rise to synchrotron radio emission spanning a wide range of luminosities \citep{Zauderer2011,Alexander2020,DeColleLu2020, Cendes2024}. Radio observations provide one of the most direct probes of this interaction \citep{GianniosMetzger2011}, allowing the energetics, velocity, geometry, and environment of the outflow to be constrained through the evolution of the radio light curve \citep{Berger2012,DeColle2012a, Zauderer2013,Alexander2016,vanVelzen2016,Alexander2017,Cendes2021SwiftJ1644,Goodwin2022,Goodwin2023b,Goodwin2023a,Cendes2024}.

The interpretation of these radio observations has relied almost exclusively on the framework developed for gamma-ray burst (GRBs) afterglows, in which the explosion is approximated as an impulsive release of energy that rapidly approaches the self-similar Blandford--McKee solution before decelerating through interaction with the surrounding medium \citep{BlandfordMcKee1976,SariPiranNarayan1998,GranotSari2002}. This approximation is well justified for GRBs because the central engine deposits nearly all of its energy before the blast wave undergoes significant deceleration \citep{MR1997,KPS1997}.

Relativistic TDEs differ in an important respect. The energy injection associated with the fallback of stellar debris naturally extends over weeks to months \citep{Rees1988,EK1989,RR2009,KrolikPiran2011,KrolikPiran2012,GuillochonRamirezRuiz2013,LS2020}, and modeling of observed TDE light curves suggests that this energy release commonly persists for several months, and in some events for years, well beyond the epoch of peak luminosity \citep{Mockler2021}. Whether prolonged energy injection modifies the blast-wave evolution depends on the competition between the duration of energy injection and the deceleration time of the outflow, the latter being determined by the properties of the circumnuclear medium \citep{BlandfordMcKee1976}. Although previous numerical studies have investigated the dynamics and radio emission from relativistic TDE jets, \citep{DeColle2012TDE,GianniosMetzger2011,MetzgerGianniosMimica2012,BarniolDuranPiran2013,Mimica2015,Generozov2017}, most studies have adopted the impulsive approximation, although \citet{DeColle2012TDE} included sustained energy injection. The regime in which sustained energy injection alters the hydrodynamic evolution and the resulting radio emission has not been systematically explored.

In this Letter, we investigate the evolution of relativistic TDE blast waves with sustained energy injection. Using relativistic hydrodynamic simulations, we compare continuously powered outflows with impulsive explosions having the same total energy, compute the resulting synchrotron radio emission, and extend the calculations to two dimensions to quantify the effects of jet geometry and viewing angle. We show that, for TDE-like engine durations and circumnuclear densities, the impulsive regime is generally not reached during the observationally relevant phase of the radio evolution. Instead, the blast wave remains continuously powered for a dynamically important fraction of its lifetime, leading to lower radio luminosities and substantially reduced relativistic beaming. This effect is especially severe for off-axis observers, whose light curves are much fainter than predicted by standard impulsive models. As a result, radio constraints on the production rate and beaming fraction of relativistic TDE jets \citep{Generozov2017} that rely on impulsive afterglow models may need to be revisited.

\begin{figure}
\centering
\includegraphics[width=0.45\textwidth]{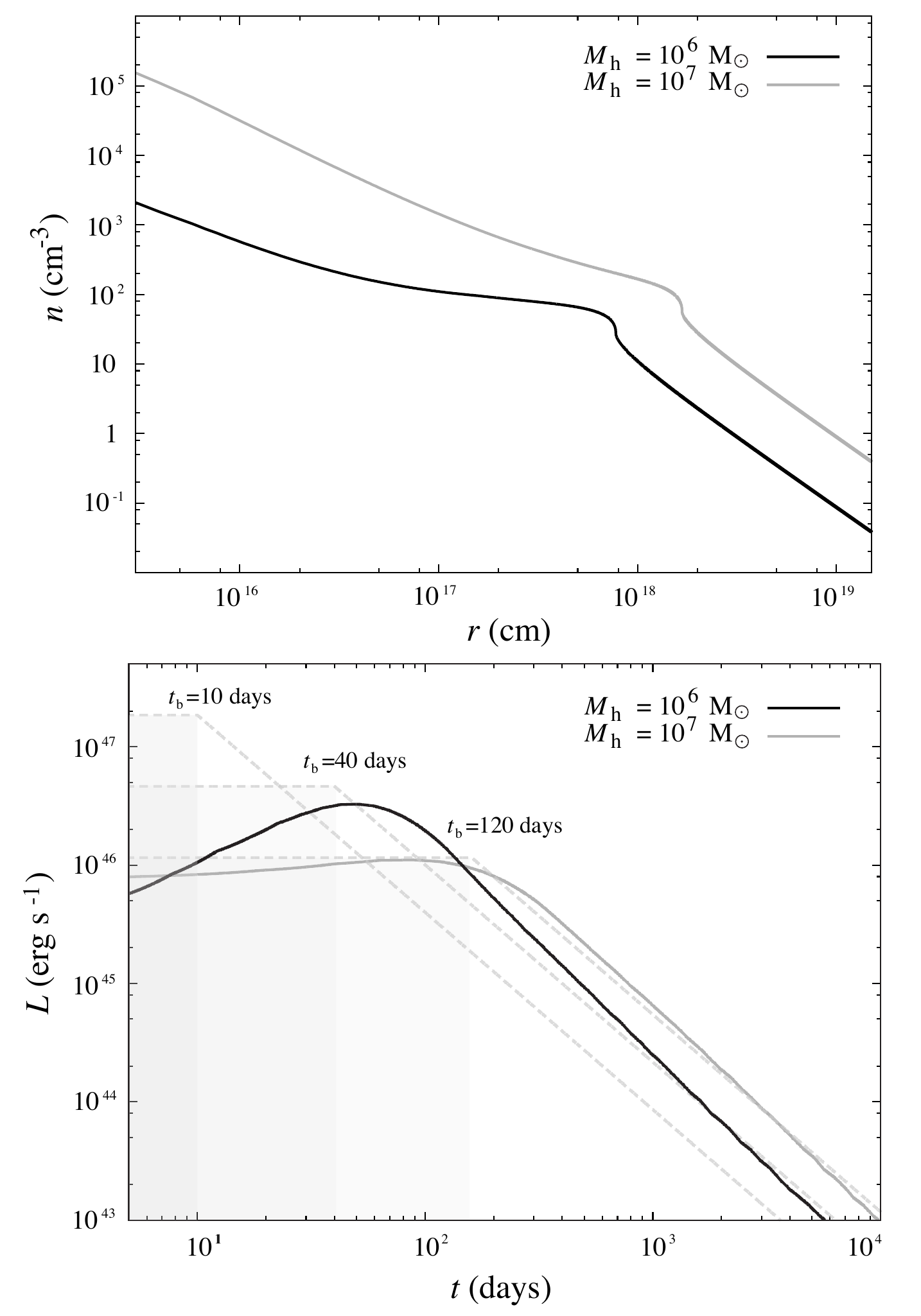}
\caption{\emph{Top:} Density stratification adopted for the circumnuclear medium surrounding supermassive black holes with masses $M_{\rm BH}=10^6$ and $10^7\,M_\odot$, based on the models of \citet{DeColle2012TDE}. The complex radial structure motivates the use of numerical simulations to follow the blast-wave evolution. \emph{Bottom:} Jet kinetic luminosity histories adopted in this work. The solid curves correspond to luminosities inferred from simulations of the disruption of a $1\,M_\odot$ star by black holes with masses $10^6$ and $10^7\,M_\odot$, while the dashed curves represent idealized models with constant luminosity for $t<t_b$ ($t_b=10$, 40, and 120 days), followed by the canonical fallback decay $L_{\rm j}\propto t^{-5/3}$. All models are normalized to the same isotropic-equivalent energy, $E_{\rm iso}=4\times10^{53}$ erg, allowing the effects of the energy injection history to be isolated.}
\label{fig1}
\end{figure}

\section{Physical Model and Numerical Methods}

\subsection{Physical Model}

We investigate the evolution of blast waves produced by relativistic TDEs by comparing continuously powered outflows with impulsive explosions of the same total energy. The calculations are motivated by the prolonged energy injection expected from fallback accretion, and are designed to determine under what conditions the standard impulsive approximation remains valid.

The external medium is taken to represent the circumnuclear environment surrounding quiescent supermassive black holes \citep{Quataert2004}. As shown in the upper panel of Figure~\ref{fig1}, the density structure is based on the stellar-wind models of \citet{DeColle2012TDE} and follows a stratified profile that transitions from a shallow inner distribution to an approximately $r^{-2}$ profile at larger radii. Owing to this complex density structure, the evolution of the blast wave cannot be described by a simple self-similar solution, requiring numerical calculations even in spherical symmetry.

The lower panel of Figure~\ref{fig1} summarizes the energy injection histories adopted in this work. Our fiducial models use jet luminosities obtained from simulations of the disruption of a $1\, M_\odot$ star by black holes with masses $10^6$ and $10^7\,M_\odot$ \citep{DeColle2012TDE,GuillochonRamirezRuiz2013}. To isolate the role of the engine duration, we also consider idealized models in which the jet luminosity remains constant for a time $t_b=10$, 20, 40, 80, 120, 160, or 200 days before declining as $t^{-5/3}$. All models are normalized to the same isotropic-equivalent energy, $E_{\rm iso}=4\times10^{53}$ erg, so that differences in the subsequent blast-wave evolution arise solely from the duration and temporal history of the energy injection.

\subsection{Hydrodynamic Simulations}
The hydrodynamic evolution is computed using the adaptive mesh refinement code \emph{Mezcal}, which solves the equations of special relativistic hydrodynamics \citep{DeColle2012a}. The code has been extensively applied to relativistic outflows associated with gamma-ray bursts and tidal disruption events \citep[e.g.,][]{DeColle2012TDE,DeColle2012b,MB2014,DeColle2018a,decolle2018b,Urrutia2021,MB2021,DeColle2022,Urrutia2023}.

The computational domain extends from $R_{\rm in}=3\times10^{15}$ cm to $R_{\rm ext}=4\times10^{19}$ cm and is evolved for $t_{\rm fin}=2R_{\rm ext}/c\approx42$ yr using a $\gamma$-law equation of state with adiabatic index $\gamma=4/3$. We perform three classes of simulations. The first consists of one-dimensional impulsive blast waves initialized using the Blandford--McKee self-similar solution with isotropic-equivalent energy $E_{\rm iso}=4\times10^{53}$ erg. The second comprises one-dimensional continuously powered models in which matter is injected through the inner boundary with Lorentz factor $\Gamma=10$ and a luminosity following the histories shown in Figure~\ref{fig1}. Finally, we perform two-dimensional simulations with jet half-opening angles $\theta_{\rm jet}=0.1$, 0.2, and 0.5 rad to investigate the effects of jet collimation and viewing angle. Additional simulations explore variations in the circumnuclear density and jet opening angle while keeping the injected energy fixed.

\subsection{Synchrotron Radiation Calculations}

Radio light curves are calculated in post-processing by assuming that electrons accelerated at the shock front follow a power-law distribution, $n(\gamma)\propto\gamma^{-p}$. A fraction $\epsilon_e$ of the post-shock internal energy is transferred to non-thermal electrons, while a fraction $\epsilon_B$ is converted into magnetic-field energy \citep{SariPiranNarayan1998}. Assuming the emitting region is optically thin, the synchrotron emissivity is computed in every computational cell using the local fluid properties \citep{DeColle2012a}. In all the simulations, we employ $\epsilon_e=0.1$ and $\epsilon_B=10^{-3}$. 

Observer light curves are constructed by combining emission from approximately one thousand simulation snapshots. For each computational cell, we determine the corresponding observer arrival time from its position and velocity, and sum the contributions from all cells arriving simultaneously at the observer. A detailed description of the radiation algorithm is presented by \citet{DeColle2012a}.

The present calculations do not include synchrotron self-absorption. Consequently, the predicted radio fluxes at early times should be regarded as upper limits. We do not, however, expect synchrotron self-absorption to alter the principal trends identified in this work, such as the delayed evolution of fallback-powered jets relative to impulsive explosions or the strong suppression of off-axis emission. Incorporating synchrotron self-absorption will enable direct comparisons with the full observed radio population and will be the subject of future work.

\begin{figure}
\centering
\includegraphics[width=0.45\textwidth]{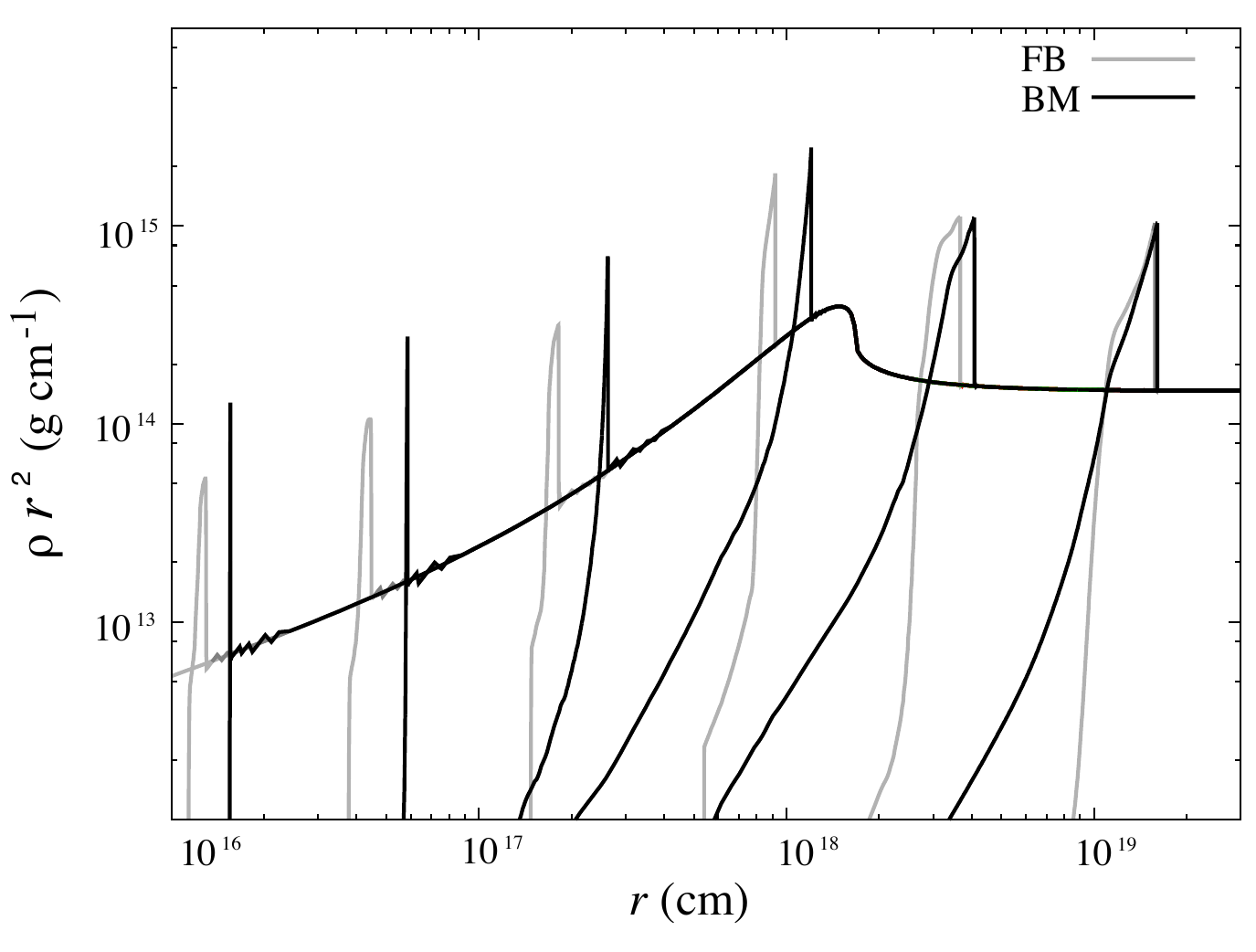}
\caption{
Evolution of the radial density profiles for one-dimensional continuously powered (gray) and impulsive (black) blast waves with the same isotropic-equivalent energy, $E_{\rm iso}=4\times10^{53}$ erg. Profiles are shown at $t=6$, 22.7, 103, 691, 4618, and 30,902 days. The powered model (FB) is driven by the fallback-powered luminosity history corresponding to a $10^7\,M_\odot$ black hole (Figure~\ref{fig1}), while the impulsive model (BM) is initialized using the Blandford--McKee self-similar solution \citep{BlandfordMcKee1976}. At early times, continuous energy injection produces a broader shocked region and prevents the formation of the thin shell characteristic of the impulsive solution. As the blast wave accumulates its final energy, the two solutions gradually converge, although the powered blast wave retains a systematically lower shock velocity throughout the energy injection phase.
}
\label{fig2}
\end{figure}

\section{One-dimensional Dynamics}\label{sec:dyn}

The evolution of an impulsive blast wave is governed by a simple sequence of phases. After an initial period of free expansion, the ejecta begin to decelerate as they sweep up the surrounding medium, eventually approaching the Blandford--McKee self-similar solution \citep{BlandfordMcKee1976}. This evolution assumes that essentially all of the explosion energy is deposited before significant deceleration occurs. In contrast, a continuously powered outflow receives its energy progressively, so that the forward shock is driven by only a fraction of its eventual energy at any given time. Consequently, two blast waves with the same final energy can evolve very differently. In this section, we investigate how sustained energy injection modifies the hydrodynamic evolution relative to the standard impulsive solution.

Figure~\ref{fig2} compares the density evolution of continuously powered and impulsive blast waves having identical isotropic-equivalent energies. The impulsive model is initialized using the Blandford--McKee solution, whereas the powered model is supplied with energy following the luminosity histories shown in Figure~\ref{fig1}. At early times, the powered solution develops a broader shocked region than its impulsive counterpart because fresh material continues to be injected behind the forward shock, continuously transferring energy and momentum into the shocked ejecta. In contrast, the impulsive blast wave rapidly settles into the familiar thin-shell structure characteristic of self-similar evolution \citep{BlandfordMcKee1976}.

The structural differences between the two solutions are a direct consequence of the different energy injection histories. As the blast wave accumulates an increasing fraction of its final energy, its evolution gradually approaches that of an impulsive explosion with the same total energy. During the energy injection phase, however, the forward shock propagates systematically more slowly because only a fraction of the eventual explosion energy has been deposited into the flow. Consequently, the continuously powered blast wave remains less energetic than its impulsive counterpart at any given time, despite both models converging to the same final energy.

\begin{figure}
\centering
\includegraphics[width=8cm]{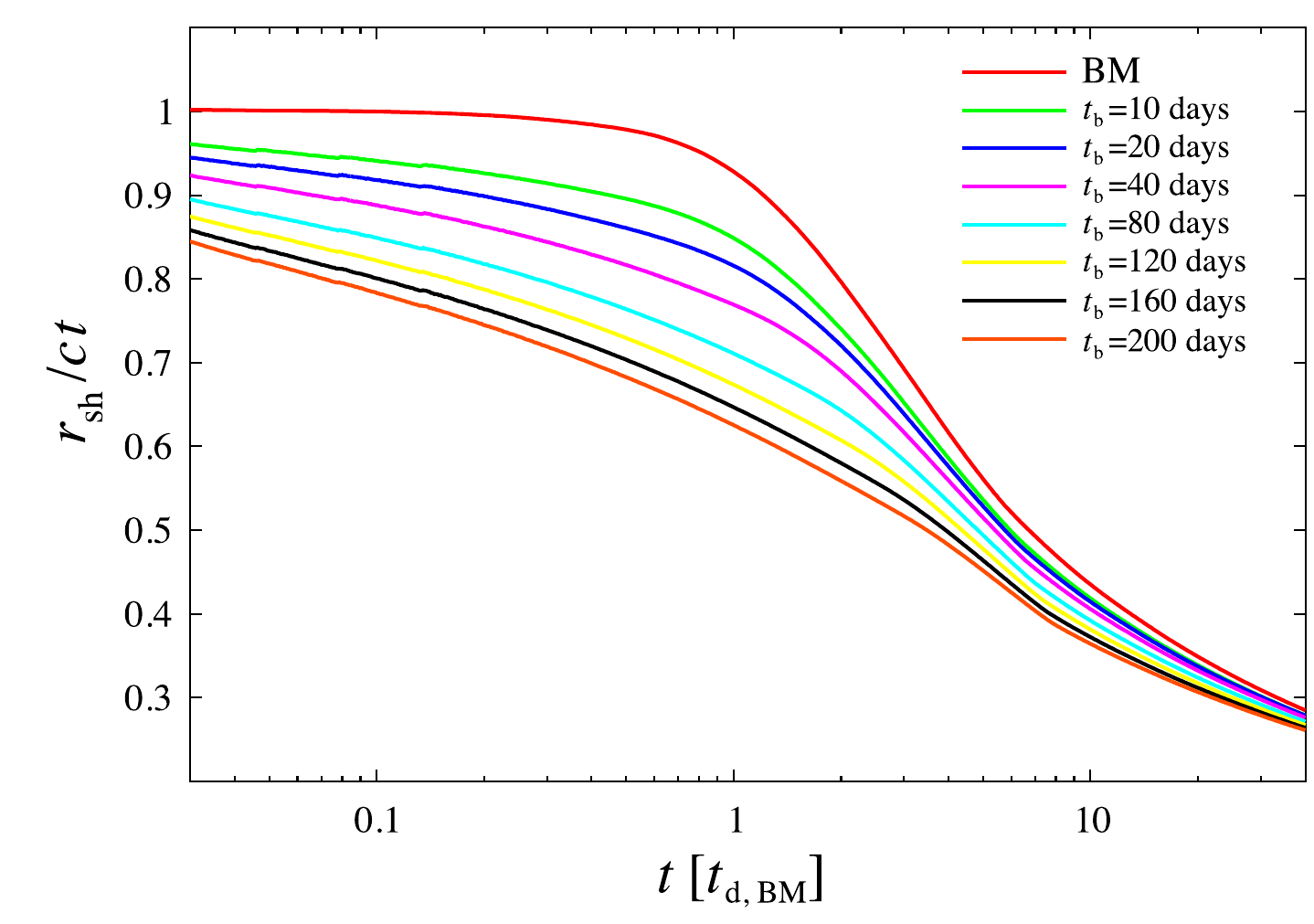}
\caption{
Evolution of the mean forward shock velocity, $\langle \beta_{\rm sh}\rangle = r_{\rm sh}/ct$, for one-dimensional impulsive and continuously powered blast waves with the same isotropic-equivalent energy, $E_{\rm iso}=4\times10^{53}$ erg. The impulsive model (BM) is initialized using the Blandford--McKee self-similar solution, while the powered models assume constant energy injection for $t<t_b$, followed by the canonical fallback decay $L_{\rm j}\propto t^{-5/3}$. The circumnuclear medium  corresponds to a $10^7\,M_\odot$ black hole (Figure~\ref{fig1}). Increasing the duration of energy injection systematically reduces the shock velocity because, at any given time, only a fraction of the eventual explosion energy has been transferred to the blast wave. Once the bulk of the energy has been injected, the powered solutions gradually converge toward the impulsive evolution.
}
\label{fig3}
\end{figure}

To quantify the global expansion of the blast wave, we define the mean forward shock velocity as $\langle \beta_{\rm sh}\rangle \equiv r_{\rm sh}/ct$, where $r_{\rm sh}$ is the forward shock radius. This quantity measures the average propagation speed of the blast wave up to a given time and provides a convenient diagnostic for comparing models with different energy injection histories.  Figure~\ref{fig3} shows the evolution of $\langle \beta_{\rm sh}\rangle$ for models with different energy injection times, $t_b$. The impulsive blast wave provides the limiting case, reaching the highest expansion velocity immediately after launch before gradually decelerating as it sweeps up the ambient medium. In contrast, continuously powered models exhibit systematically lower shock velocities throughout the energy injection phase. The reduction becomes progressively more pronounced as the duration of energy injection increases, reflecting the fact that a smaller fraction of the total explosion energy has been transferred to the blast wave at any given time.

Once the bulk of the energy has been injected, the shock velocity gradually converges toward the impulsive solution. The rate of this convergence depends on the duration of the energy injection relative to the deceleration timescale of an equivalent impulsive blast wave \citep[e.g.,][]{Ramirez-Ruiz2001}. As we show in the following section, this reduction in shock velocity has important observational consequences, leading to systematically lower radio luminosities and substantially weaker relativistic beaming than predicted by impulsive blast-wave models.

\section{Radio Emission from Powered Blast Waves}\label{sec:radio}

The reduced shock velocities found in the previous section directly influence the synchrotron emission, since both the post-shock magnetic field and the energy imparted to relativistic electrons depend on the strength of the forward shock \citep{SariPiranNarayan1998,GranotSari2002}. We now investigate how sustained energy injection modifies the radio light curves and identify the physical parameter that determines when the standard impulsive approximation remains valid.

\begin{figure}
\centering
\includegraphics[width=8cm]{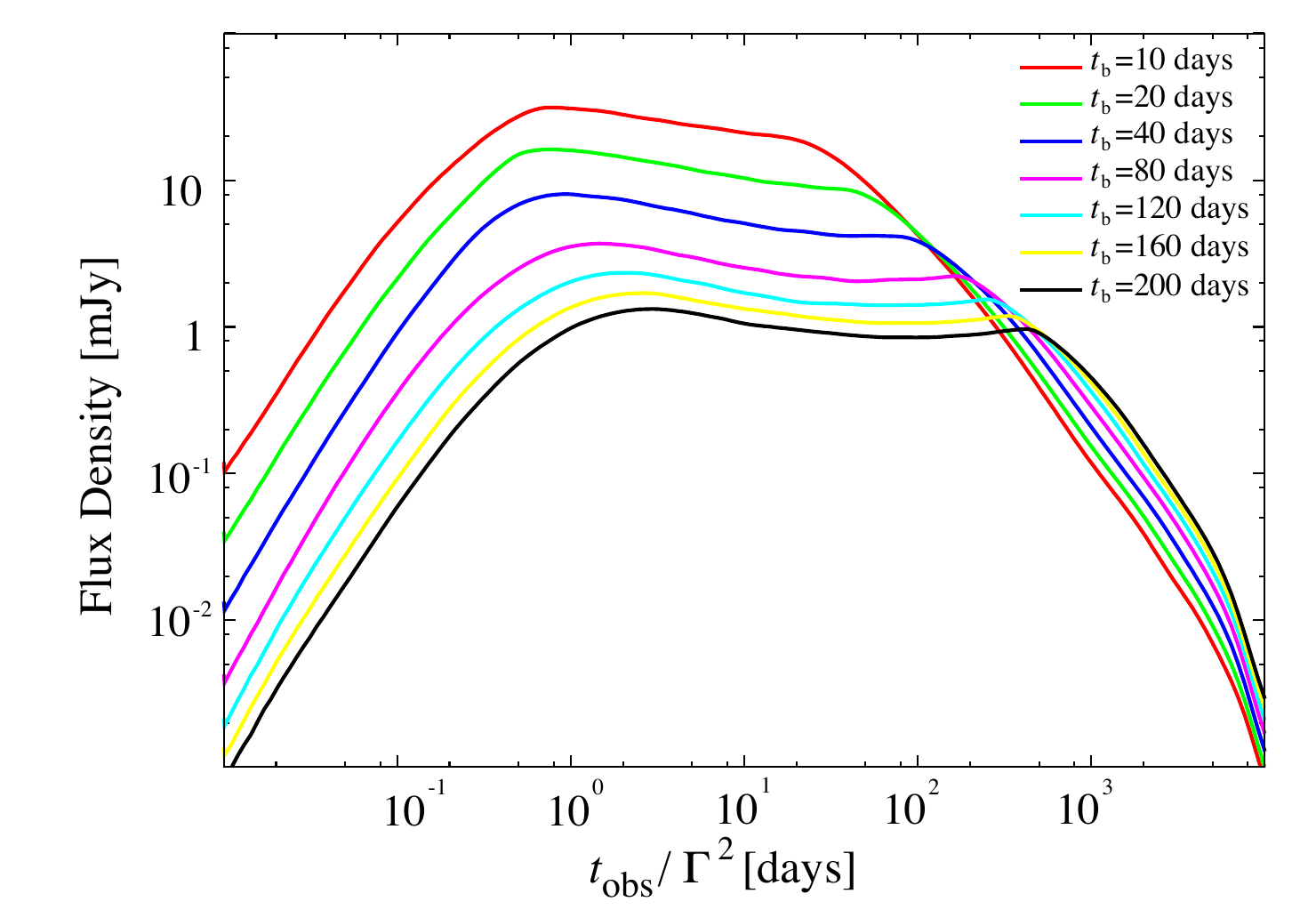}
\caption{Radio light curves at $\nu=4.86$ GHz for the constantly-powered, one-dimensional simulations shown in Figure~\ref{fig3}. The colored curves correspond to continuously powered blast waves with different energy injection times, $t_b$.
Increasing the duration of energy injection systematically lowers the peak radio luminosity and delays the time of maximum emission.
}
\label{fig4}
\end{figure}

Figure~\ref{fig4} compares the radio light curves of continuously powered 
blast waves having the same isotropic-equivalent energy. Despite their identical final total energies, the powered models are systematically fainter during the period of active energy injection. As shown in Section~\ref{sec:dyn}, the forward shock propagates more slowly. The reduced shock velocity lowers the post-shock energy density and the amount of swept-up material participating in the emission, leading to weaker magnetic fields and a smaller population of accelerated electrons. This suppresses the synchrotron emissivity and produces systematically fainter radio emission.

The differences become increasingly pronounced as the duration of energy injection increases. Models with longer engine lifetimes not only reach lower peak luminosities but also peak at later times, reflecting the delayed transfer of energy to the blast wave. Only after most of the available energy has been injected do the powered solutions gradually approach the evolution of an impulsive explosion with the same total energy.

These results demonstrate that interpreting radio observations using impulsive blast-wave models can systematically overestimate the luminosity expected from continuously powered outflows. The magnitude of this discrepancy depends on the relative importance of continued energy injection and blast-wave deceleration, a relationship that we now quantify.

\begin{figure}
\centering
\includegraphics[width=8cm]{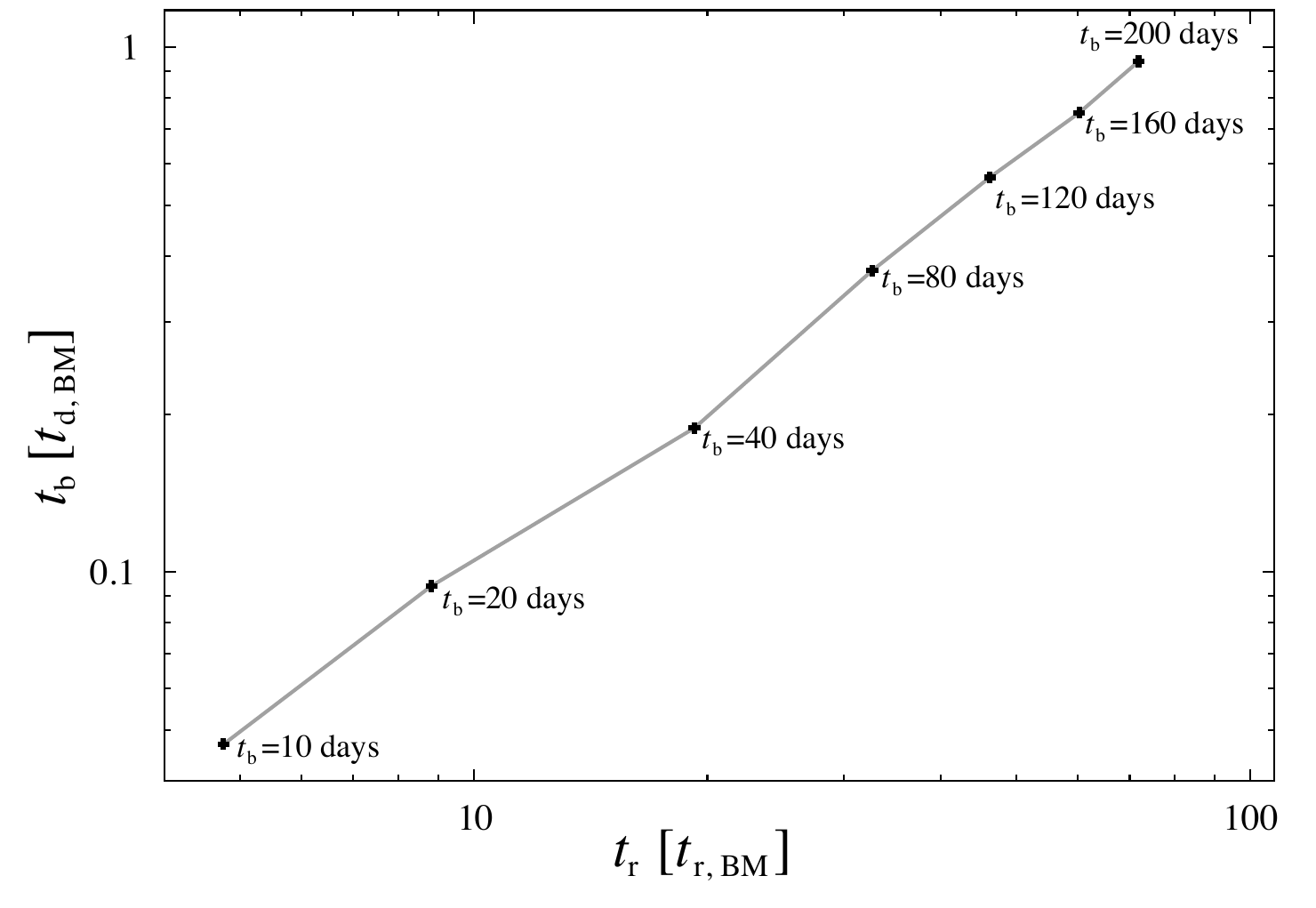}
\caption{
Energy injection time, $t_b$, normalized to the deceleration time of an equivalent impulsive Blandford--McKee blast wave, $t_{\rm d,BM}$, as a function of the radio peak time, $t_r$, normalized to the peak time of the corresponding impulsive solution, $t_{\rm r,BM}$ for the one-dimensional simulations shown in Figures~\ref{fig3} and \ref{fig4}. Although the energy injection terminates before the Blandford--McKee deceleration time for all models ($t_b<t_{\rm d,BM}$), the radio emission peaks substantially later than in the impulsive case. 
}
\label{fig5}
\end{figure}

The characteristic deceleration time of an equivalent impulsive Blandford--McKee blast wave, $t_{\rm d,BM}$, provides the natural dynamical timescale against which the duration of energy injection should be compared. One might therefore expect that once the engine shuts off before significant deceleration occurs ($t_b<t_{\rm d,BM}$), the blast wave would rapidly converge toward the impulsive solution and produce nearly identical radio emission. Our simulations show that this expectation is overly simplistic.

Figure~\ref{fig5} compares the duration of energy injection, normalized to the Blandford--McKee deceleration time ($t_b/t_{\rm d,BM}$), with the time of the radio peak, normalized to that of the corresponding impulsive solution ($t_r/t_{\rm r,BM}$). Although all of the models considered here satisfy $t_b\lesssim t_{\rm d,BM}$, their radio emission peaks substantially later than in the impulsive case. The delay increases systematically with the duration of energy injection, demonstrating that the blast wave continues to retain the imprint of its injection history well after the central engine has ceased operating.

These results show that the deceleration time alone does not determine when the impulsive approximation becomes valid. Instead, the radio emission reflects the cumulative hydrodynamic response of the blast wave to prolonged energy injection. The quantities $t_b$ and $t_{\rm d,BM}$ nevertheless remain the relevant characteristic timescales, as their ratio provides a natural measure of the departure from the impulsive limit. Our simulations indicate that the Blandford--McKee approximation provides a good description only when the duration of energy injection is less than approximately ten percent of the impulsive deceleration time, $t_b/t_{\rm d,BM}\lesssim0.1$. For larger ratios, the radio peak is progressively delayed relative to the impulsive solution, demonstrating that the blast wave retains a significant memory of its energy injection history.

\subsection{Robustness to Realistic Circumnuclear Environments}

The results presented above were obtained using idealized energy injection histories, allowing the effects of sustained powering to be isolated. An important question is whether the same conclusions apply to the more realistic fallback-powered jets and circumnuclear density profiles expected in tidal disruption events \citep{Quataert2004}.

\begin{figure}
\centering
\includegraphics[width=8cm]{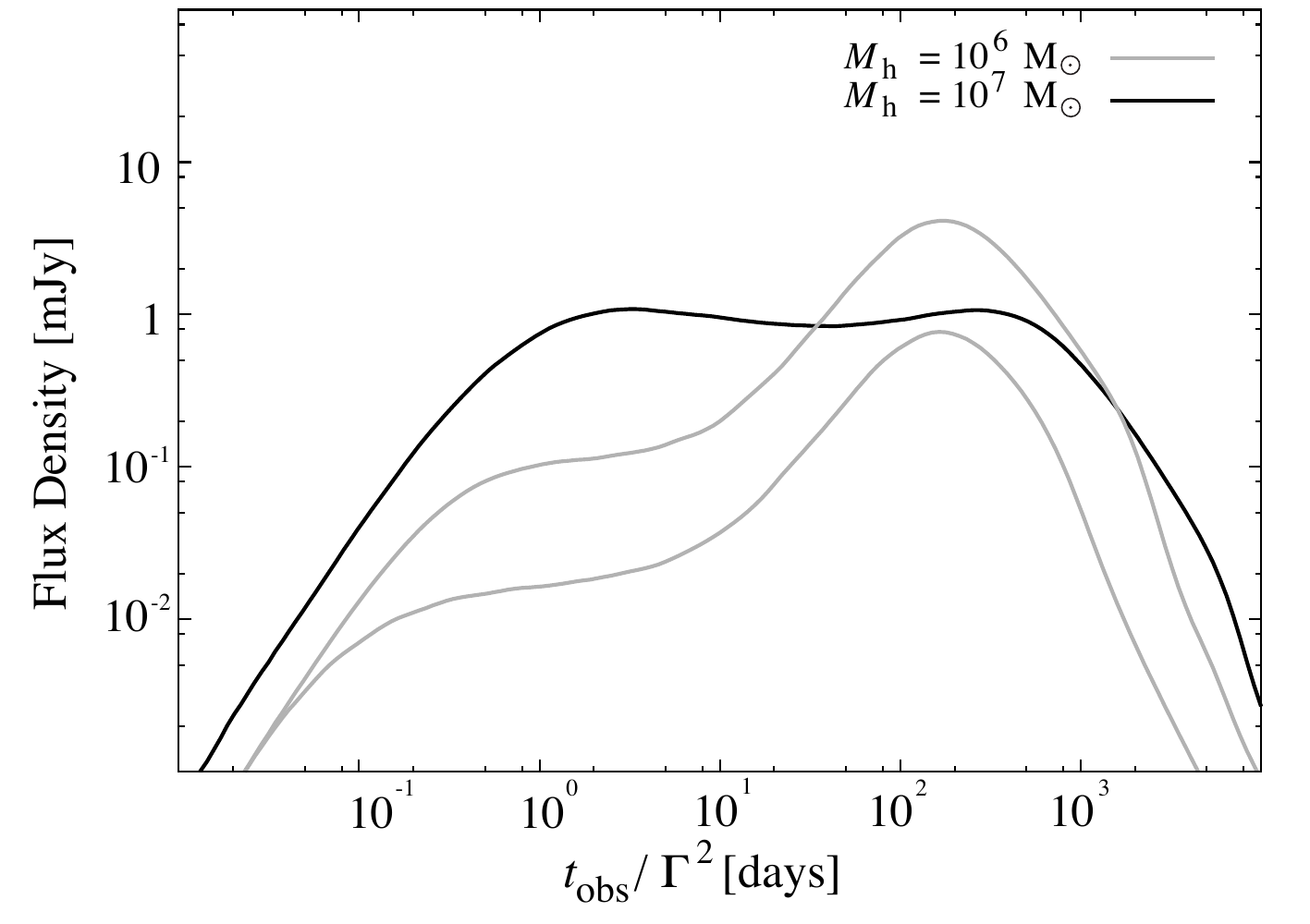}
\caption{
Radio light curves at $\nu=4.86$ GHz for one-dimensional simulations using the fallback-powered jet luminosities and circumnuclear density profiles shown in Figure~\ref{fig1} for $10^6\,M_\odot$ and $10^7\,M_\odot$ black holes. The black curve corresponds to the fiducial $10^7\,M_\odot$ model. The gray curves correspond to models with a $10^6\,M_\odot$ black hole that adopt the fiducial density profile shown in Figure~\ref{fig1} and an ambient density profile enhanced by a factor of ten. Although the external density modifies the deceleration time and peak luminosity, the delayed radio evolution produced by sustained energy injection persists across the full range of environments considered.
}
\label{fig6}
\end{figure}

Figure~\ref{fig6} addresses this question by repeating the calculations using the physically motivated density profiles and fallback accretion histories shown in Figure~\ref{fig1}. In addition to the fiducial model, we consider external density normalizations that differ by an order of magnitude, thereby spanning the range of environments expected around quiescent supermassive black holes \citep[e.g.,][]{Alexander2020}.

As expected, increasing the ambient density reduces the deceleration time and increases the radio luminosity \citep{DeColle2012TDE}. Nevertheless, the fallback-powered energy injection remains dynamically important, and the radio light curves continue to peak significantly later than the corresponding impulsive solutions. This remains true even when the ambient density is reduced by an order of magnitude relative to the fiducial profile, indicating that the delayed radio evolution is robust over a broad range of circumnuclear densities.

These calculations demonstrate that the conclusions reached in the previous subsection are not an artifact of the idealized constant-luminosity models. Instead, they persist for realistic tidal disruption events, where the characteristic fallback timescale remains dynamically important over a broad range of circumnuclear densities. Thus, the prolonged energy injection expected in TDEs \citep{Mockler2021} naturally produces radio evolution that departs from the standard impulsive Blandford--McKee picture, even in relatively low-density environments.

\begin{figure*}
\centering
\includegraphics[width=\textwidth]{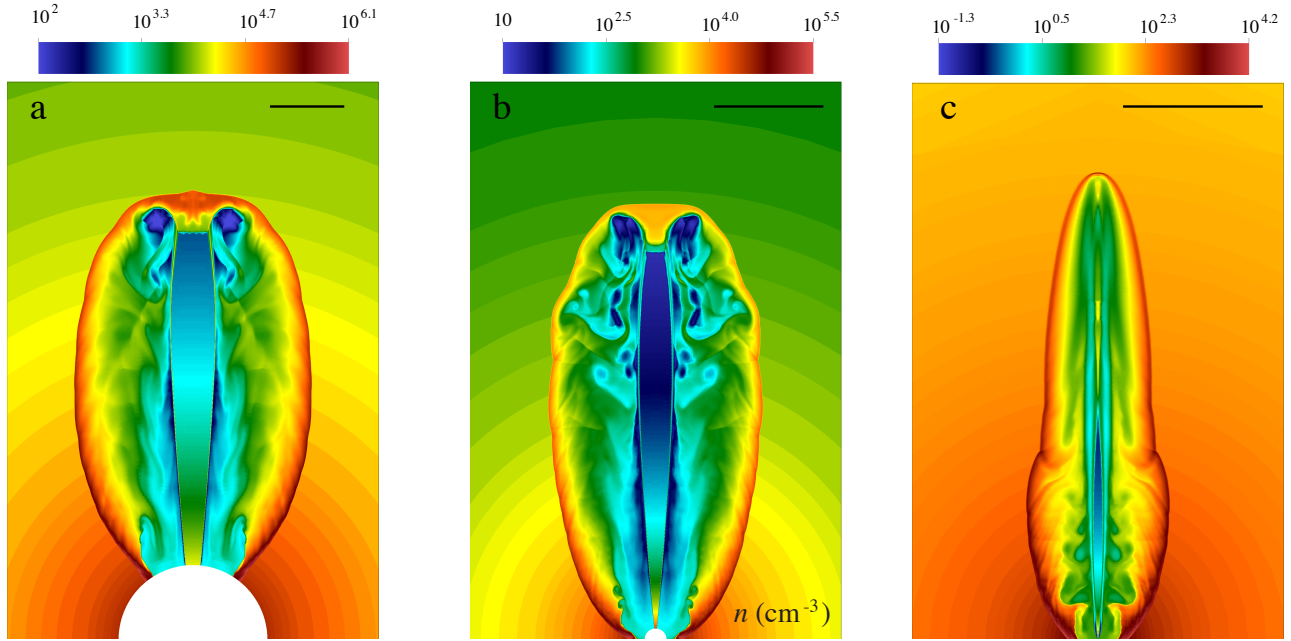}
\caption{
Density maps from a two-dimensional simulation of a relativistic jet propagating through the circumnuclear medium surrounding a $10^7\,M_\odot$ black hole at $t=7$, 47, and 380 days (left to right). The jet is injected at the inner boundary with Lorentz factor $\Gamma_{\rm jet}=10$, isotropic-equivalent energy $E_{\rm iso}=4\times10^{53}$ erg, and a luminosity history corresponding to the fallback-powered model shown in Figure~\ref{fig1}. The interaction of the jet with the ambient medium produces the characteristic forward and reverse shock structure while inflating a hot cocoon that expands laterally and confines the relativistic jet. The physical scale shown in each panel increases with time: the horizontal bars correspond to $3\times10^{15}$, $3\times10^{16}$, and $3\times10^{17}$ cm, respectively. The simulation is performed on a spherical polar grid extending from $3\times10^{15}$ to $2\times10^{18}$ cm with a jet half-opening angle $\theta_{\rm jet}=0.1$ rad. The computational grid is resolved with $200\times2$ cells in the radial and polar directions and 14 levels of adaptive mesh refinement.
}
\label{fig7}
\end{figure*}

\section{Multidimensional Evolution}

The one-dimensional calculations presented above demonstrate that sustained energy injection reduces the shock velocity and delays the hydrodynamic evolution of the blast wave. Relativistic tidal disruption events, however, produce collimated jets whose evolution cannot be fully captured by spherical models \citep{GianniosMetzger2011,Bloom2011,MetzgerGianniosMimica2012,DeColle2012TDE,Andreoni2022}. In particular, lateral expansion, cocoon formation, and the angular distribution of energy influence the observed radio emission, especially for off-axis observers \citep[e.g.,][]{Rhoads1999,Granot2002,Ramirez-Ruiz2005}.

To investigate these effects, we perform two-dimensional relativistic hydrodynamic simulations of powered jets with the same energy injection histories considered in the previous sections. These calculations connect the modified radial dynamics identified in the one-dimensional models to the angular structure that determines the observed radio emission.

Figure~\ref{fig7} illustrates the evolution of a continuously powered relativistic jet propagating through the circumnuclear medium. The jet head is composed of the familiar forward and reverse shock structure \citep{DeColle2012TDE}, separated by a contact discontinuity. Material shocked at the jet head expands sideways, inflating a hot cocoon \citep{Ramirez-Ruiz2002,Matzner2003} that confines the jet while driving a broader, slower shock into the surrounding medium. As the jet propagates through the inner circumnuclear medium, where the ambient density declines more slowly than $r^{-2}$ (Figure~\ref{fig1}), the overpressured cocoon collimates the jet through a series of internal recollimation shocks that redistribute its kinetic energy while maintaining a narrow relativistic core \citep{Bromberg2011}. At larger radii, where the density approaches an $r^{-2}$ profile, the external confinement weakens and the jet transitions toward nearly free expansion.

\begin{figure}
\centering
\includegraphics[width=0.45\textwidth]{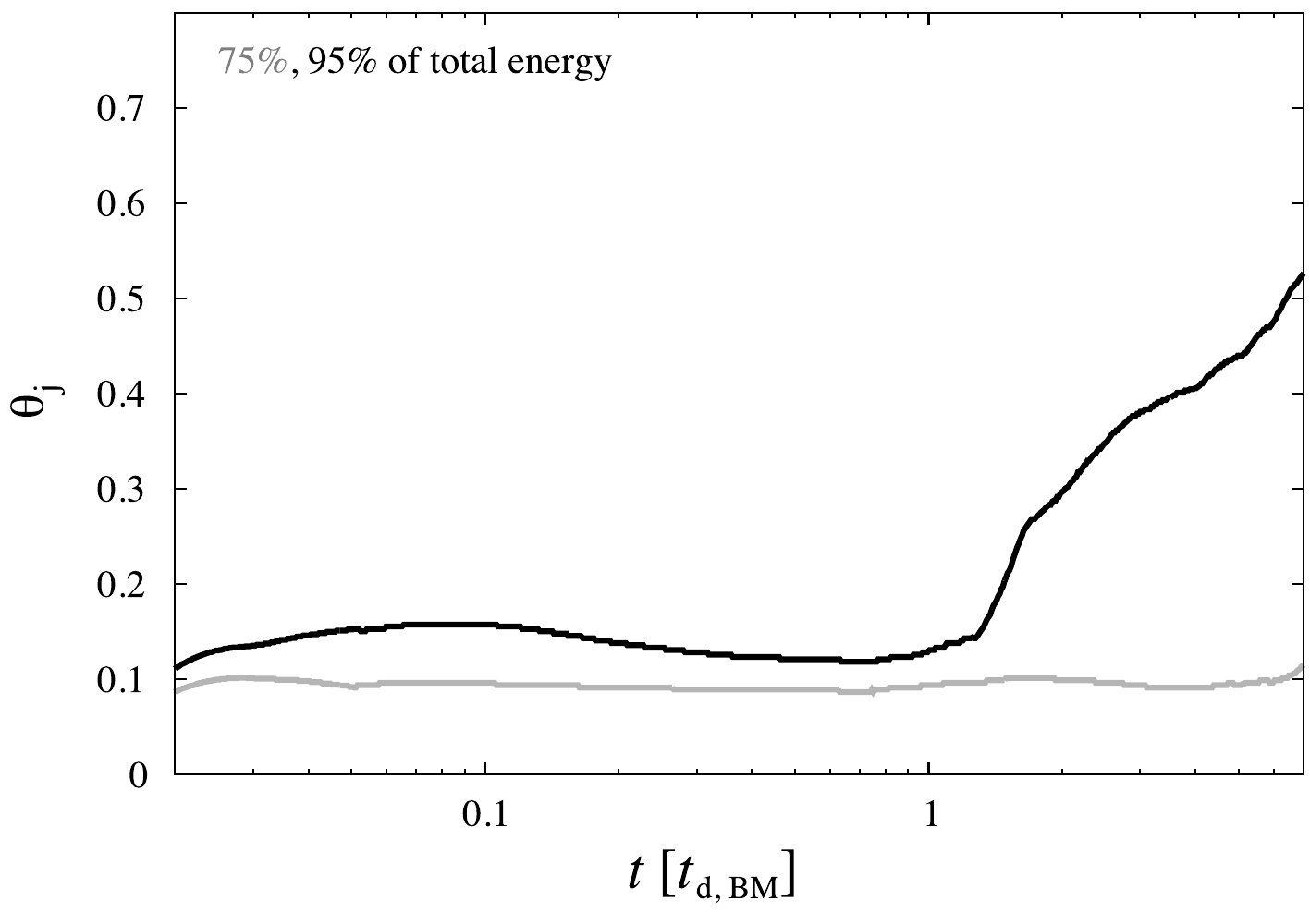}
\caption{
Evolution of the jet opening angle in the two-dimensional simulation shown in Figure~\ref{fig7}. The opening angle is defined as the angle enclosing 75\% and 95\% of the total isotropic-equivalent jet energy ($E_{\rm iso}=4\times10^{53}$ erg). The opening angle containing 75\% of the jet energy remains nearly constant throughout most of the evolution, demonstrating that the relativistic core undergoes very little lateral expansion. In contrast, the opening angle enclosing 95\% of the energy gradually increases on timescales comparable to the Blandford--McKee deceleration time, $t_{\rm d,BM}$, reflecting the lateral expansion of the lower-energy material surrounding the jet core.
}
\label{fig8}
\end{figure}

Figure~\ref{fig8} quantifies the lateral evolution of the jet by measuring the opening angles enclosing 75\% and 95\% of the total jet energy. The relativistic core, which contains most of the explosion energy, remains remarkably well collimated throughout the evolution, exhibiting very little lateral expansion. In contrast, the lower-energy material surrounding the core spreads gradually, producing only a modest increase in the opening angle enclosing 95\% of the total energy. Thus, despite the continued growth of the cocoon, the bulk of the explosion energy remains concentrated within a narrow angular region.

The persistence of a narrow relativistic core has important observational consequences. Because the observed emission from relativistic jets depends sensitively on Doppler beaming, the modified angular structure and reduced shock velocities identified here are expected to alter both the brightness and detectability of off-axis radio emission. The following section quantifies these effects by comparing the radio emission produced by powered and impulsive jets for observers located both within and outside the initial jet opening angle.

\section{On-axis and Off-axis Radio Emission}
The multidimensional calculations demonstrate that sustained energy injection modifies both the radial and angular evolution of relativistic TDE jets. We now investigate how these hydrodynamic differences translate into observable radio emission. In particular, we compare the light curves predicted by continuously powered jets with those of impulsive explosions having the same total energy, considering observers both within and outside the initial jet opening angle.

\begin{figure}
\centering
\includegraphics[width=0.45\textwidth]{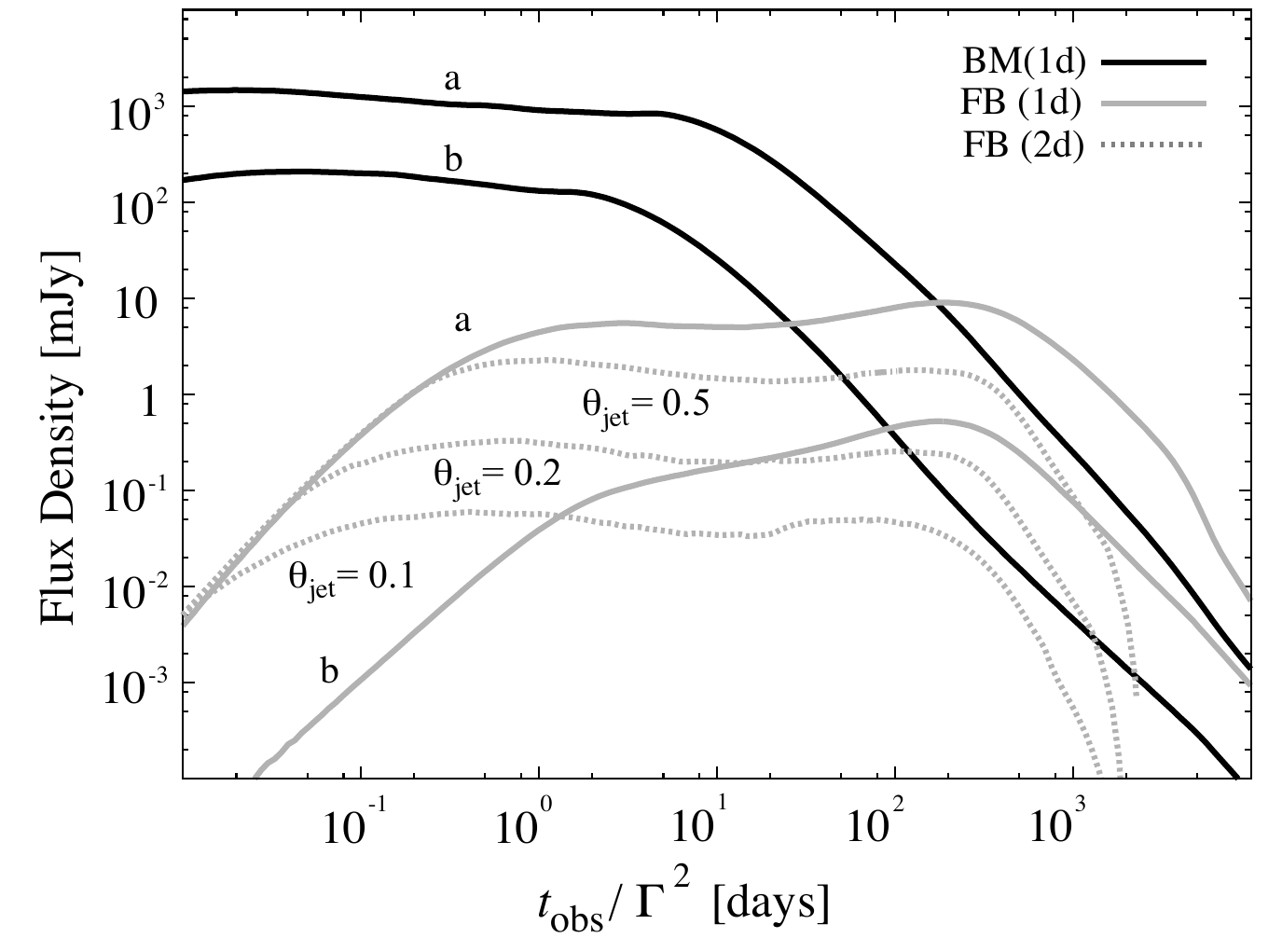}
\caption{
On-axis radio light curves at $\nu=4.86$ GHz comparing impulsive Blandford--McKee (BM) blast waves with fallback-powered (FB) jet models. The FB simulations adopt the physically motivated jet luminosity histories shown in Figure~\ref{fig1}. Results are shown for isotropic-equivalent energies of $E_{\rm iso}=2\times10^{51}$ erg (models labeled ``b'') and $E_{\rm iso}=4\times10^{53}$ erg (models labeled ``a''). The two-dimensional calculations assume jet half-opening angles of $\theta_{\rm jet}=0.1$, 0.2, and 0.5 rad. Relative to the BM solutions, the FB models exhibit systematically later and fainter radio peaks as a consequence of sustained energy injection. For the FB models, the two-dimensional jets produce lower on-axis fluxes than the corresponding one-dimensional calculations because the emitting surface is reduced in collimated outflows, with narrower jets yielding progressively lower radio luminosities.
}
\label{fig9}
\end{figure}

Figure~\ref{fig9} compares the radio light curves computed from the one- and two-dimensional simulations for observers located along the jet axis. The excellent agreement between the two calculations demonstrates that the one-dimensional models capture the radial dynamics responsible for the on-axis emission. The small differences that develop at late times arise from the modest lateral expansion of the cocoon but have little effect on the overall light-curve evolution. Consequently, the conclusions derived from the one-dimensional calculations regarding the effects of prolonged energy injection remain valid for realistic collimated jets when viewed on axis.

The situation changes dramatically for observers located outside the initial jet opening angle. Off-axis emission depends sensitively on the degree of relativistic beaming, which is modified by the prolonged energy injection discussed above.

\begin{figure}
\centering
\includegraphics[width=0.4\textwidth]{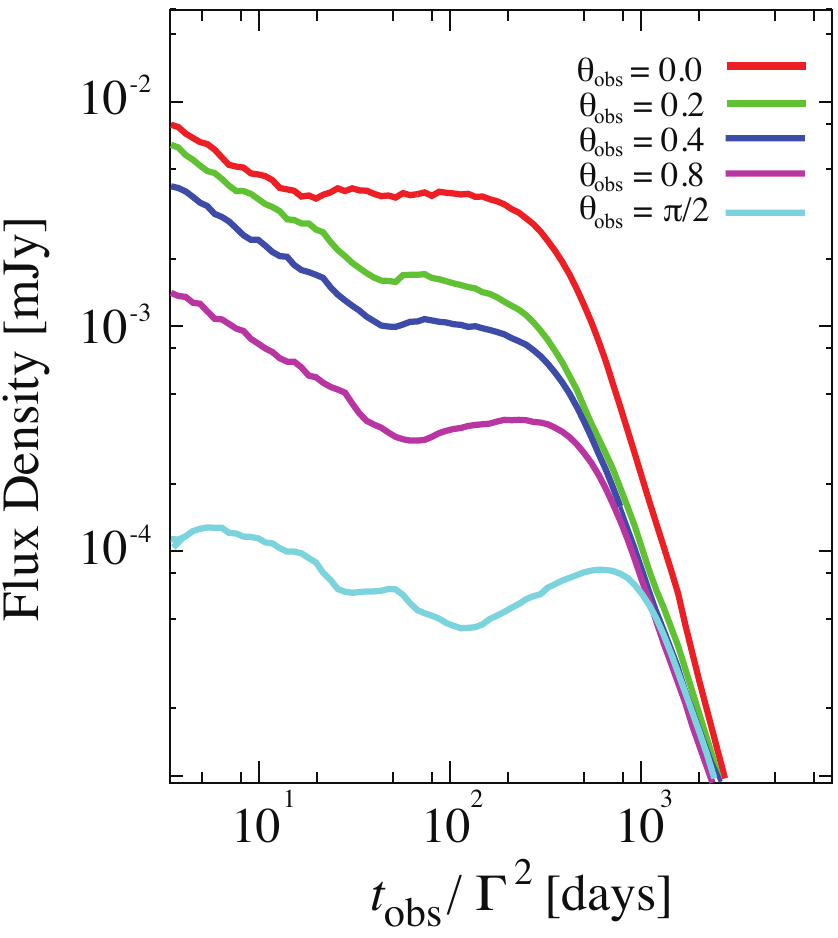}
\caption{
Radio light curves at $\nu=4.86$ GHz for a fallback-powered jet viewed from observer angles $\theta_{\rm obs}=0$, 0.2, 0.4, 0.8, and $\pi/2$. The jet follows the fallback luminosity history shown in Figure~\ref{fig1}, has an initial half-opening angle $\theta_{\rm jet}=0.1$, and an isotropic-equivalent energy $E_{\rm iso}=4\times10^{53}$ erg. As the viewing angle increases, the observed flux decreases and the radio peak shifts to later times as relativistic beaming becomes progressively weaker. The suppression of the radio emission for off-axis observers is substantially stronger than predicted by impulsive models because the lower shock velocities produced by sustained energy injection reduce the degree of Doppler boosting.
}
\label{fig10}
\end{figure}

Figure~\ref{fig10} compares the radio light curves predicted for several viewing angles. Relative to impulsive explosions with the same total energy \citep{Generozov2017,DeColleLu2020}, continuously powered jets produce systematically fainter off-axis emission over the entire evolution. The difference becomes increasingly pronounced with viewing angle.

This suppression results from two closely related effects. First, the lower shock velocities produced by sustained energy injection reduce the Lorentz factor of the emitting material, weakening relativistic Doppler boosting. Second, the limited lateral expansion demonstrated in Figure~\ref{fig8} prevents a substantial fraction of the jet energy from being redistributed toward larger angles. As a result, powered jets remain highly concentrated around the jet axis while producing significantly less emission for off-axis observers than predicted by impulsive models.

These results have important consequences for interpreting radio observations of tidal disruption events. Because standard afterglow models generally assume impulsive energy injection, they are expected to overpredict the off-axis radio luminosity of fallback-powered relativistic jets. Consequently, non-detections of radio emission place weaker constraints on the presence of relativistic jets than previously inferred, implying that a larger fraction of TDEs may harbor successfully launched jets whose emission remains undetected because of viewing-angle effects.

\section{Discussion}\label{sec:discussion}

The standard framework used to interpret radio emission from relativistic tidal disruption events assumes that the outflow is powered by an impulsive release of energy and subsequently evolves as a decelerating Blandford--McKee blast wave. This approximation has provided the foundation for modeling radio afterglows over the past decade. The calculations presented here demonstrate that this assumption is generally not satisfied for relativistic TDEs. Because the jet is powered by fallback accretion, energy continues to be supplied to the blast wave over timescales that are often comparable to or longer than its deceleration time. As a result, the hydrodynamic evolution retains memory of the central engine throughout much of the observable radio evolution.

The consequences of this sustained energy injection are systematic. Compared with an impulsive explosion of the same total energy, fallback-powered jets drive slower forward shocks, produce weaker post-shock magnetic fields, and generate lower synchrotron luminosities. Although these effects are already evident for on-axis observers, they become considerably more pronounced away from the jet axis. The combination of reduced shock velocities and the persistence of a narrow relativistic core substantially weakens Doppler boosting, producing radio light curves that are significantly fainter than predicted by impulsive models.

These results suggest that the observed diversity of radio tidal disruption events does not necessarily require a corresponding diversity of outflow mechanisms. We propose that a single physical picture, in which relativistic jets are continuously powered by fallback accretion, could naturally explains the broad range of radio luminosities through variations in viewing angle, circumnuclear density, black hole mass, and the duration of energy injection. Rather than representing fundamentally different classes of explosions, some of the diversity observed in radio TDEs may instead reflect the continuous coupling between the evolving central engine and the expanding blast wave.

\begin{figure}
\centering
\includegraphics[width=0.45\textwidth]{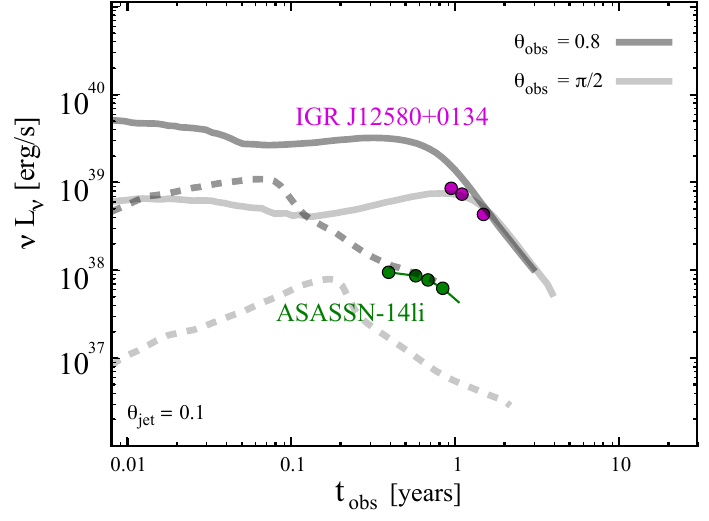}
\caption{
Comparison between the radio light curves predicted by the fallback-powered jet model and the observed radio luminosities of the early sample of radio-detected tidal disruption events. The models adopt the fallback luminosity history shown in Figure~\ref{fig1} for a $10^7\,M_\odot$ black hole and an initial jet half-opening angle of $\theta_{\rm jet}=0.1$. Solid curves correspond to the fiducial circumnuclear density profile, while dashed curves show models with an ambient density lower by an order of magnitude. The observational data are shown for comparison \citep{vanVelzen2016, Alexander2016,Alexander2017}.
}
\label{fig11}
\end{figure}

Figure~\ref{fig11} provides an illustrative comparison between our calculations and the early sample of radio-detected tidal disruption events that have frequently been interpreted as sub-energetic, mildly relativistic outflows. Although these events are substantially fainter than the prototypical relativistic jet Swift~J1644+57 \citep{Zauderer2011}, our calculations demonstrate that fallback-powered relativistic jets observed away from the jet axis naturally populate a similar range of radio luminosities. This comparison raises the possibility that at least some events previously modeled as slow outflows may instead be consistent with successful relativistic jets viewed off axis.

The comparison shown in Figure~\ref{fig11} is particularly relevant for the early sample of radio-detected tidal disruption events, which were typically monitored only sporadically and primarily at epochs when the emission had already become optically thin. Consequently, the present calculations provide a physically motivated first assessment of whether fallback-powered relativistic jets can reproduce the observed range of radio luminosities. A definitive comparison with the expanding TDE sample will require extending the radiation calculations to include synchrotron self-absorption and performing a broader suite of multidimensional simulations spanning the relevant parameter space. We defer this population-level analysis to future work.

Such calculations are particularly timely given the rapid growth of the radio TDE sample and the recent synthesis presented by \citet{Alexander2026}. Incorporating sustained energy injection into models of relativistic TDE jets provides a physically motivated framework for interpreting this expanding population and may ultimately revise the inferred fraction of tidal disruption events that successfully launch relativistic jets.

We note that our model is restricted to the prompt radio afterglow produced by the initial jet launched during the tidal disruption event. We do not attempt to model the delayed radio emission observed in a growing number of TDEs months to years after disruption \citep{Cendes2022,Cendes2024,Alexander2026}, which has been suggested to originate from delayed accretion \citep[e.g.,][]{Cannizzo1990}, accretion-state transitions \citep[e.g.,][]{DeColle2012TDE}, or renewed outflow activity \citep[e.g.,][]{Cendes2022} rather than from the early relativistic ejecta.

Our results suggest that the radio properties of relativistic tidal disruption events cannot, in general, be understood independently of the long-term evolution of the central engine. Incorporating sustained fallback-powered energy injection into models of TDE jets provides a new framework for interpreting the growing radio sample and may substantially revise our understanding of how frequently tidal disruption events successfully launch relativistic jets.

\section*{Acknowledgments}
We are grateful for insightful conversations with K. Alexander, K. Auchettl, Y. Cendes, J. Dai, J. Lynch, N. Velez, and B. Mockler.
We acknowledge the computing time granted by DGTIC UNAM on the supercomputer Miztli (project LANCAD-UNAM-DGTIC-281). 
We acknowledge support from the DGAPA/PAPIIT grant IN113424. The UC Santa Cruz team is supported in part by the Heising-Simons Foundation, the Vera Rubin Presidential Chair at UC Santa Cruz, and the National Science Foundation through grants AST-2307710, AST-2206243, and AST-1911206. This research was also supported by the Lamat Institute \citep{Quinteros2025}.

\bibliography{sample701}{}
\bibliographystyle{aasjournalv7}

\end{document}